# One nanometer thin carbon nanosheets with tunable conductivity and stiffness

**Andrey Turchanin[1*], André Beyer[1], Christoph Nottbohm[1], Xianghui Zhang[1], Rainer Stosch[2], Alla Sologubenko[3], Joachim Mayer[3], Peter Hinze[2], Thomas Weimann[2], Armin Gölzhäuser[1]**

*[1]Fakultät für Physik, Universität Bielefeld, 33615 Bielefeld, Germany*

*[2]Physikalisch-Technische Bundesanstalt, 38116 Braunschweig, Germany*

*[3]Gemeinschaftslabor für Elektronenmikroskopie, RWTH Aachen, 52074 Aachen, Germany*

E-mail: turchanin@physik.uni-bielefeld.de

Tel.: +49-521-1065376

Fax: +49-521-1066002

**Abstract:**

We present a new route for the fabrication of ultrathin (~1 nm) carbon films and membranes, whose electrical behavior can be tuned from insulating to conducting. Self-assembled monolayers of biphenyls are cross-linked by electrons, detached from the surfaces and subsequently pyrolized. Above 1000K, the cross-linked aromatic monolayer forms a mechanically stable graphitic phase. The transition is accompanied by a drop of the sheet resistivity from ~$10^8$ to ~$10^2$ kΩ/sq and a mechanical stiffening of the nanomembranes from ~10 to ~50 GPa. The technical applicability of the nanosheets is demonstrated by incorporating them into a microscopic pressure sensor.

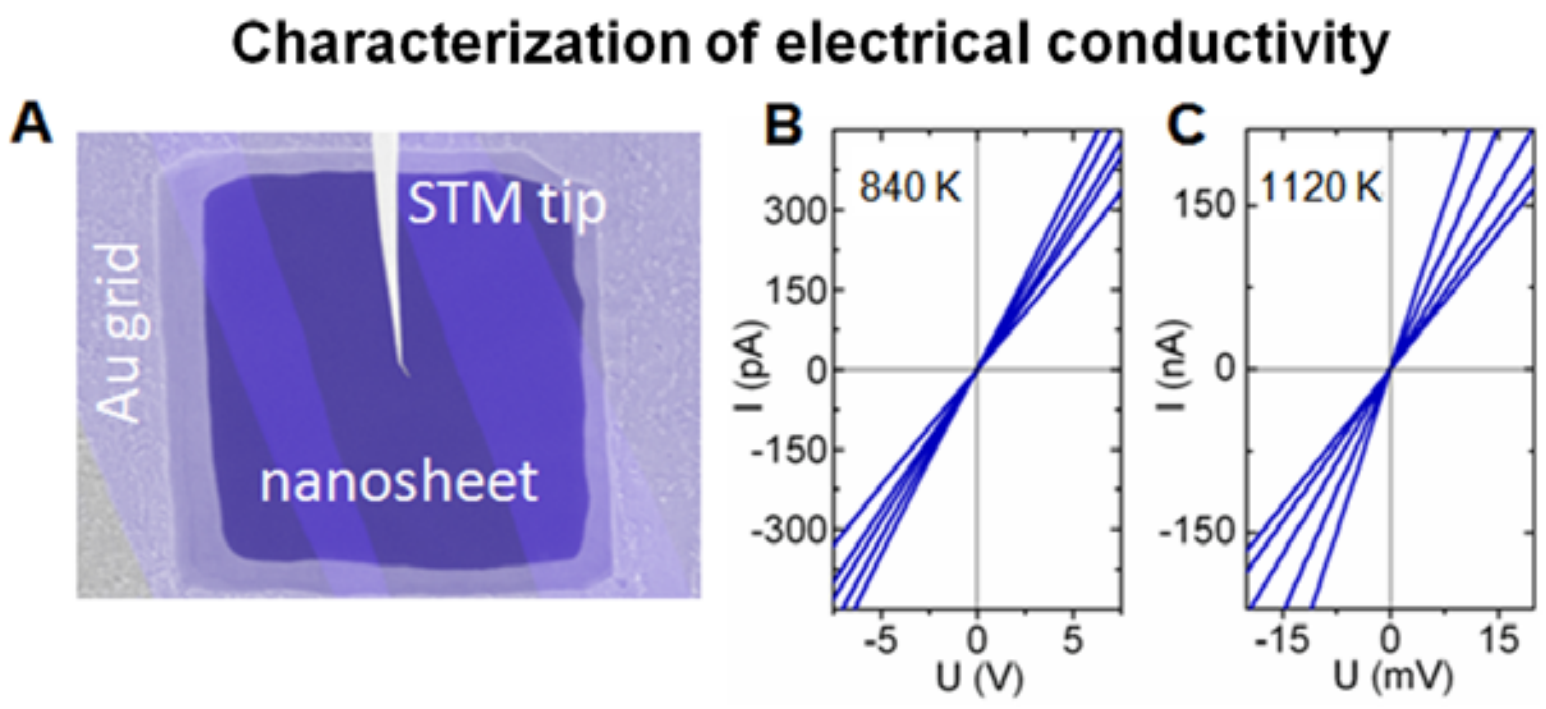

For centuries carbon allotropes have been identified with three-dimensional carbon phases like diamond and graphite. In the last few decades, nanoscopic zero-dimensional (fullerenes[1]) and one-dimensional (carbon nanotubes[2]) carbon allotropes generated a manifold of research due to their potential use in nanotechnology. The recent discovery of a two-dimensional carbon allotrope – graphene[3] – marks an important breakthrough in physics, since it has long been argued that free-standing atomically thin materials cannot exist at ambient conditions. The subsequent aim for novel applications of two-dimensional carbon ignited significant research efforts[4-13]. For example, it is highly desirable to have atomically thin carbon sheets with tunable electrical, mechanical, and optical properties as well as with controllable size, shape and chemical functionality. Nanoscale electronics[9], nanoelectromechanical systems (NEMS)[10], as well as nano- and biosensors[14] could particularly benefit from the incorporation of such two-dimensional carbon sheets in composite materials and devices[9]. However, methods currently used for graphene fabrication such as mechanical exfoliation of highly oriented pyrolytic graphite[3], epitaxial methods[6, 12], or reduction of graphene oxide[8], can only partly fulfill these demands. Thus, there is a great need for novel paths to two-dimensional carbon allotropes.

Highly oriented pyrolytic graphite (HOPG) is the best ordered artificially made three-dimensional graphitic allotrope. It can be fabricated via pyrolysis of bulk aromatic polymers in the temperature range from 1000K to 3000K[15]. In analogy to this bulk transformation, we suggest that the pyrolysis of a molecular thin film of aromatic molecules is a promising path for the generation of two-dimensional carbon. Biphenylthiols form densely packed self-assembled monolayers (SAMs) with a thickness of ~1 nm on gold surfaces[16]. Such aromatic SAMs could act as suitable precursors in a pyrolytic reaction. However, due to the low thermal stability of thiolates[17], they desorb from the surface at temperatures that are much lower (350-450K) than those required for pyrolysis. We have recently found that the temperature induced desorption of biphenylthiols on gold is inhibited[18], when biphenyls were cross-linked by electron irradiation in a molecular sheet[19]. In this report we show that vacuum pyrolysis can transform ~1 nm thick aromatic molecular nanosheets from an insulating to a conducting state. The resulting carbon nanosheets are atomically thin and mechanically stable as suspended membranes even at temperatures above 1200 K and their resistivity and stiffness are determined by the annealing temperature.

To prepare carbon nanosheets (details in supporting online material (SOM)) biphenyl molecules are self-assembled on a substrate from solution and subsequently cross-linked by electron irradiation, Fig. 1A. Both size and shape of the nanosheets are determined by this initial exposure. Modern electron beam lithography and exposure tools allow the fabrication of sheets from macroscopic ($cm^2$) down to nanometer sizes and in arbitrary shapes[20]. The nanosheets are then lifted from their surface and transferred to another solid substrate or holey structures, such as transmission electron microscope (TEM) grids, where the nanosheets become suspended free-standing nanomembranes. The transfer itself is very simple. First, a polymeric transfer medium ("glue") is applied to the sheet and the original substrate is dissolved. The hardened glue with the attached nanosheet is then placed onto another solid surface or a TEM grid and finally, the glue is dissolved (see SOM), leaving the nanosheet on the new substrate. Fig. 1B shows an optical micrograph of the section of a large (~5 $cm^2$) carbon nanosheet that has been transferred from gold onto a silicon wafer with a ~300 nm thick silicon oxide layer. The color variation between the bare part of the silicon oxide surface and the part covered by nanosheet allows for visualization of the cross-linked biphenyl monolayer[21]. Within the nanosheet some dark lines are clearly visible. These lines are folds in the sheet that occurred during the transfer process. The thickness of the nanosheet has been determined by X-ray photoelectron spectroscopy and atomic force microscopy (AFM) to be ~1.2 nm (SOM), which is in good agreement with the height of a biphenyl molecule. Fig. 1C shows ~10 µm wide nanosheet lines that were written by electron beam lithography and then transferred onto silicon with a ~300 nm oxide layer. Compared to the $cm^2$ sized sheet in Fig. 1B, the 10 µm wide nanosheet lines show almost no folds, indicating that small sheets have a lower tendency to wrinkle during the transfer process.

Fig. 1D shows a scanning electron micrograph (SEM) of a nanosheet that has been transferred onto a TEM grid with 130x130 $\mu m^2$ squared holes. The two holes on the left are covered by an intact homogeneous nanomembrane. In the upper right hole the membrane shows some folds, and in the lower right hole, the nanomembrane has ruptured. Nanomembranes on TEM grids with small holes (<10 µm) show very few such rupture defects, however, yield decreases with increasing hole size.

We pyrolysed nanomembranes on TEM grids at temperatures from ~800K to ~1300K. Fig. 1E shows a TEM micrograph of a cross-linked biphenyl nanosheet on a gold grid that has been annealed at ~1100 K in ultra high vacuum (UHV). An intact nanosheet (with a few folds) that spans an 11x11 $\mu m^2$ hole is clearly seen in Fig. 1E. Scanning Auger microscopy revealed that this suspended membrane consists only of carbon (SOM). This temperature stability is quite remarkable for a macroscopically large membrane with a thickness of only ~1 nm.

In the next step, we explored the electrical properties of the heated nanosheets in suspended (membranes) and supported states (films). Fig. 2 shows the sheet resistivity as a function of the annealing temperature. The resistivity was determined at room temperature after the respective annealing steps. Nanosheets suspended on a gold grid were contacted by the tip of a scanning tunneling microscope; resistance was then determined by a two-point measurement in UHV. Fig. 2A shows a scanning electron micrograph of a tungsten tip touching a nanosheet that suspends over an 11x11 $\mu m^2$ squared opening. Additional resistivity measurements were carried out under ambient conditions. To this end, nanosheets heated directly on gold substrate in UHV where transferred on silicon oxide and their sheet resistance was determined under ambient conditions by a four-point measurement (SOM). The sheet resistivity values measured in UHV and in ambient are in a very good agreement. A measurable electrical current is detected after annealing at ~800 K. Here the sheet resistivity corresponds to ~$10^8$ kΩ/sq. Upon annealing to temperatures between 800 and 1200K, we find linear current/voltage curves (Fig. 2 B,C,E). Increasing the annealing temperature to ~1200 K, drops the sheet resistivity to ~100 kΩ/sq, which demonstrates the clear metallic nature of the film. This resistivity is only one order of magnitude higher than that of a defect free graphene monolayer[4], and ~100 times lower than the sheet resistivity of single chemically reduced graphene oxide sheets[22], which are currently most favored for mass production of graphene[8].

The structural transformations that occur upon annealing in the cross-linked aromatic monolayer were investigated by Raman spectroscopy and high resolution transmission electron microscopy (HRTEM). Again, nanosheets supported on silicon oxide substrates and films suspended on TEM grids were analyzed at room temperature after annealing. For annealing temperatures above 700 K, two peaks at ~1350 and ~ 1590 $cm^{-1}$ are observed in the Raman spectrum (Fig. 3A). These bands are referred to as the so-called D- and G-peaks which are characteristic for $sp^2$-bonded, honeycomb structured carbon allotropes[23]. Their positions, shapes and the intensity ratio $I$(D)/$I$(G) provide information about the degree of order in the carbon network*[24]*. At ~730K the D-peak has its maximum intensity at 1350 $cm^{-1}$ while the G-peak has its maximum intensity at 1592 $cm^{-1}$ and shows a shoulder at 1570 $cm^{-1}$. At higher annealing temperatures, the shoulder in the G peak disappears. The band narrows and its position successively shifts to higher wave numbers reaching 1605 $cm^{-1}$ at ~1200 K. Simultaneously, the ratio $I$(D)/$I$(G) increases from ~0.75 to ~1 (Fig. 3B). The maximum of the D-peak almost remains at the same wave number while above ~950 K a shoulder appears at ~1180 $cm^{-1}$. The observed temperature dependent changes in the Raman spectra are characteristic for a phase transition from an amorphous to a nanocrystalline carbon network[24]. Considering the thickness of the carbon nanosheet (1 nm), it is reasonable to attribute these changes to the formation of a nanosize graphene network. The Raman spectra also correlate very well with the successive decrease of the sheet resistivity for increasing annealing temperatures.

The occurrence of structural ordering in the annealed sheets can be observed by HRTEM studies of suspended membranes. For non-annealed membranes, both high resolution imaging and selected area electron diffraction (SAED) show only the presence of amorphous material, Fig 3C, D.  In annealed specimens, extended areas with curvy, nearly parallel fringes indicating the presence of graphitic material were found, Fig. 3E. The areas where fringes are observed, alternate with areas where they are not present. These observations clearly indicate that distinct structural changes occur in the nanosheet upon annealing and that the intrinsic properties of this two-dimensional material must vary accordingly[25, 26]. The line profiles across the regions with fringes, Fig.3E (1,2), give a periodicity of 0.35±0.03 nm, which is close to the interplanar spacing of the close-packed planes in graphite (0.342 nm). In our experiments, the scattered intensity modulations corresponding to the 0.35±0.03 nm periodicity of the fringes could not be found in the diffraction patterns from the investigated areas. Our evaluation showed that the corresponding reciprocal distance still lies within the strong tails of the central beam of the diffraction pattern. However, the enlargement of the SAED pattern (Fig. 3F, SOM) taken from a much larger area as depicted in Fig.3E shows two distinct rings (marked 1 and 2) corresponding to the real space periodicities of 0.11±0.02 nm (ring 2) and 0.20±0.02 nm (ring 1). These can be interpreted as to correspond to the major indices (0-110) (1.23 Å spacing) and (1-210) (2.13 Å spacing) of highly in-plane oriented nanocrystalline graphitic sheets observed along the [0002] zone axis. The sharpness and intensity of the rings increase in specimens annealed at higher temperatures indicating progressing ordering in the nanomembrane.

The structural transformation of a cross-linked aromatic monolayer is also reflected in the mechanical properties. To quantify these, we fabricated a nanomechanical pressure sensor in which the nanosheet acts as a membrane, cf. Fig. 4A. This rather simple device demonstrates the utilization of carbon nanomembranes as a nanomechanical transducer. Freely suspended nanosheets were mounted onto a sealed pressure cell, and a well defined pressure difference between both sides of the membrane was applied. The resulting membrane deflection was measured by AFM and used to determine the Young's modulus and the residual strain of nanosheets by bulge tests[27]. Fig. 4B shows an AFM image of a nanosheet annealed at ~900 K without applied pressure. Although the membrane is pushed down ~15 nm by the tip, it remains intact. By applying a pressure of ~450 Pa to the sealed cell under the membrane, an upward deformation (bulging) occurs (Fig. 4C). This deformation is quantified by recording the AFM tip height at the membrane centre as function of the applied pressure.

Deformation datasets are presented in Fig. 4D which contains three successive measurement cycles. All measured data lie on one curve and no hysteresis is detectable, i.e. the deformation is elastic without any permanent change within the investigated strain range of up to 0.6 %. The absence of any hysteresis shows that the nanosheet does not slide on the silicon frame, presumably due to a sufficiently strong van der Waals interaction. Long-term stability was tested after five month, and no changes in the elastic properties could be observed. This demonstrates that the nanomembrane deflection can be utilized for pressure sensing. A model for the elastic deformation of thin membranes under tension contains two parts[27]: (i) membrane stretching leads to a cubic dependence of the pressure to the height, (ii) membrane tension at zero pressure due to residual stress results in a linear pressure-height-dependence. The experimental data fit very well to this model which is plotted in Fig. 4D. Curve fitting yields the Young's modulus and the residual strain (see SOM). Fig. 4E shows both quantities as a function of the annealing temperature. Without thermal treatment the Young's modulus is 12 GPa. This value is comparable to the Young's modulus of multi-layered molecular/metallic nanocomposite membranes[28] that are thicker by an order of magnitude. Annealing leads to a systematic increase of the modulus with rising temperature up to 48 GPa at ~1000 K. This is in good agreement with an increasing graphitization, as the Young's modulus of graphite varies from 39 GPa to 1.1 TPa[29], depending on its orientation. The formation of residual strain in the nanosheet is most likely related to structural transformations during the cross-linking process. Without annealing the nanosheet shows a residual strain of 0.8 %.

Annealing reduces the residual strain of the nanosheet to ~0.35 % above 800 K, which correlates with the onset of conductivity.

Since nanomembranes are elastic and mechanically stable at ambient conditions they can be further utilized as sensitive diaphragms in various applications. Conducting nanomembranes may act as transducers in nanoelectromechanical systems (NEMS) and open an opportunity to build highly miniaturized pressure sensors that might eventually lead to microphones with nanometer dimensions. The possibility to chemically functionalize nanosheets by chemical lithography[30] further permits their use as highly sensitive chemical sensors that change their electromechanical characteristics upon the adsorption of distinct molecules.

In conclusion, we have shown a simple method to produce ultrathin (~1 nm) conducting carbon films and membranes based on molecular self-assembly, electron irradiation and pyrolysis. Upon annealing, cross-linked aromatic monolayers undergo a transition to a mechanically stable graphitic phase. The above experiments demonstrate a plethora of applications that take advantage from the fact that size, shape and conductivity of the films and nanomembranes are easily controlled.

**Supporting Information:**

A detailed description of materials and methods, an analysis of spectroscopic and microscopic data, a description of models for evaluation of electrical and mechanical measurements.

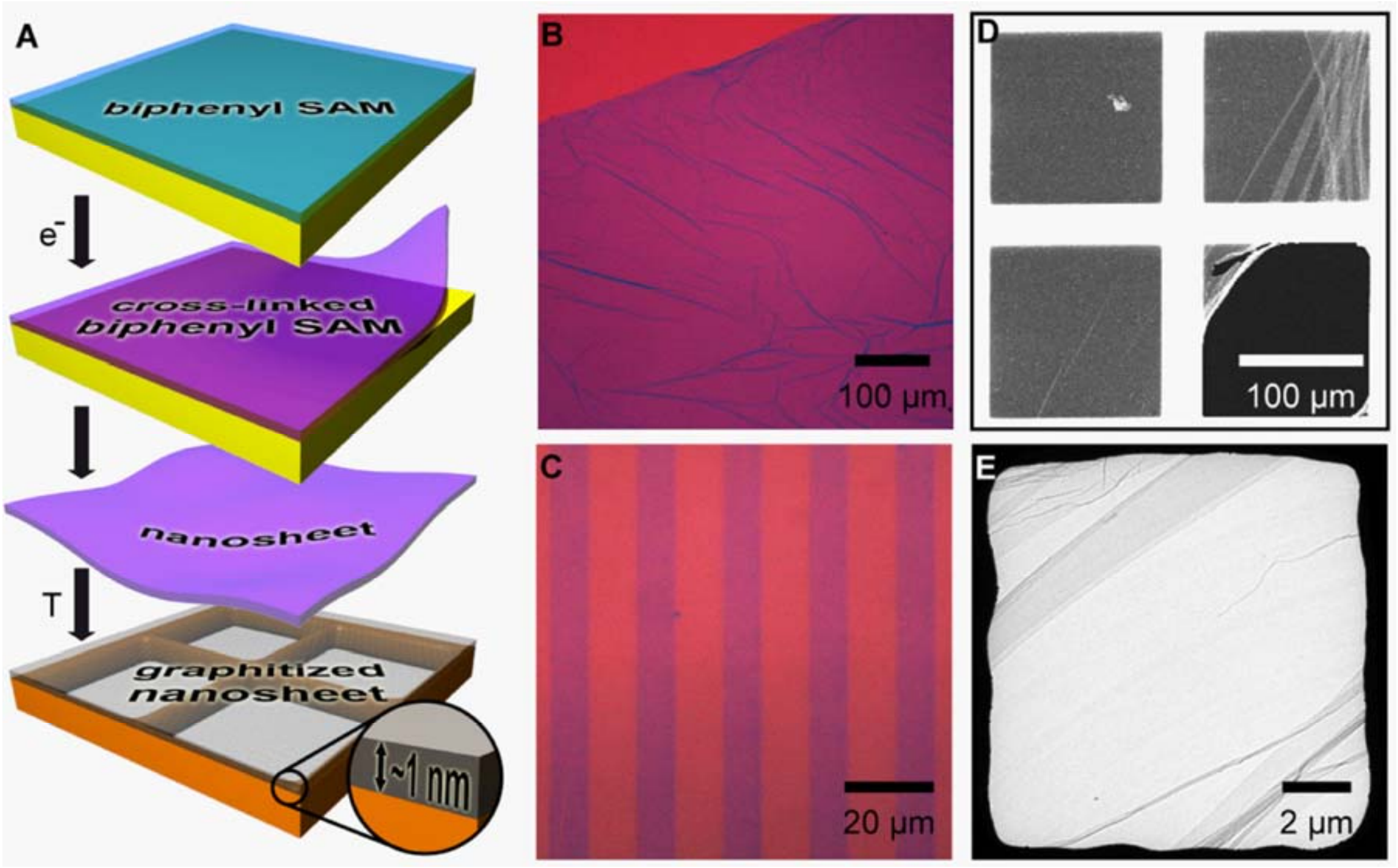

**Fig. 1. Fabrication scheme and micrographs of supported and suspended carbon nanosheets.** (A) A ~1 nm thick self-assembled monolayer (SAM) of biphenyl molecules is irradiated by electrons. This results in a mechanically stable cross-linked SAM (nanosheet) that can be removed from the substrate and transferred onto other solid surfaces. When transferred onto transmission electron microscopy (TEM) grids, nanosheets suspend over holes. Upon heating to T>1000K in vacuum (pyrolysis), nanosheets transform into a graphitic phase. (B) Optical micrograph of the section of a ~5 $cm^2$ nanosheet that was transferred from a gold surface to an oxidized silicon wafer (300 nm $SiO_2$). Some folds in the large sheet are visible that originate from wrinkling during the transfer process. (C) Optical micrograph of a line pattern of 10 µm stripes of nanosheet. The pattern was fabricated by e-beam lithography and then transferred onto oxidized silicon. Note that the small lines are almost without folds. (D) Scanning electron micrograph of four 130x130 $\mu m^2$ holes in a TEM grid after nanosheet (cross-linked biphenyl SAM) has been transferred onto the grid. Two left holes are covered by almost unfolded nanosheet. The upper right hole shows some folds, whereas in the lower right hole, the sheet has ruptured. (E) Transmission electron micrograph of a nanosheet transferred onto a TEM grid with 11x11 $\mu m^2$ holes after pyrolysis at 1100K. The hole is uniformly covered with an intact nanosheet. Some folds within the sheet are visible.

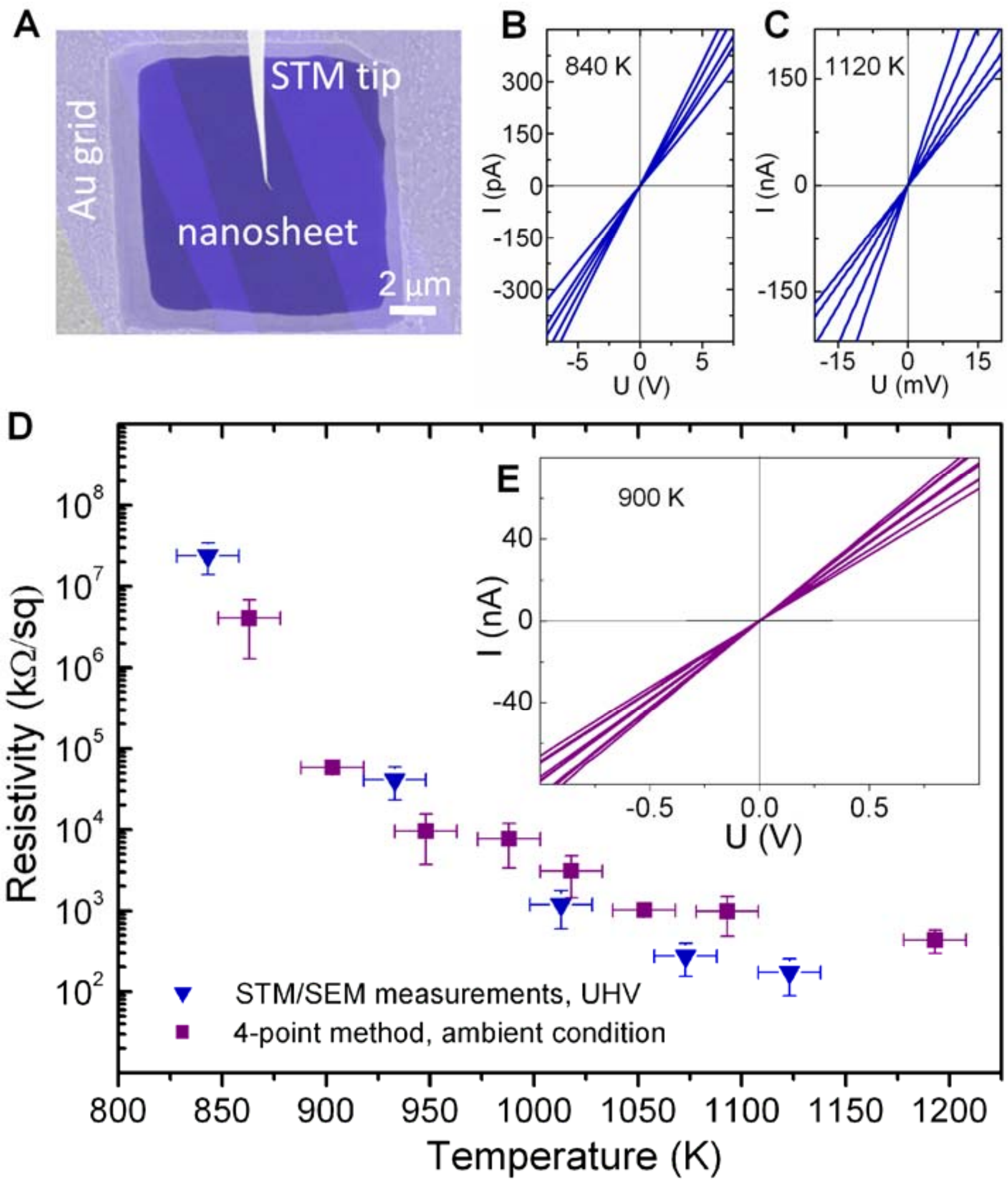

**Fig. 2. Room temperature resistivity of carbon nanosheets after annealing at different temperatures.** (A) SEM image of the tungsten STM tip establishing an electrical and mechanical contact in the centre of the annealed biphenyl nanosheet suspended on the gold grid with ~11×11 µm$^2$ squared openings, as employed for two-point resistivity measurements. A folded nanosheet (shown in false colour) was chosen for better visualisation. A boundary between the bare gold surface and the gold surface with a nanosheet can clearly be recognized in the lower left corner of the image. (B) and (C) Representative room temperature current vs. voltage data for two annealing temperatures in the two-point set-up of resistivity measurements in UHV. Each line corresponds to a measurement in a different window of the grid. (E) Room temperature current vs. voltage data (four-point set-up) of the nanosheets supported on a silicon oxide surface. The nanosheets were annealed on gold and then transferred to silicon oxide substrates. Each line corresponds to a measurement at a new position on the nanosheet. (D) Summary of the room temperature sheet resistivity as a function of annealing temperature. All samples were annealed for 30 min at the respective temperatures.

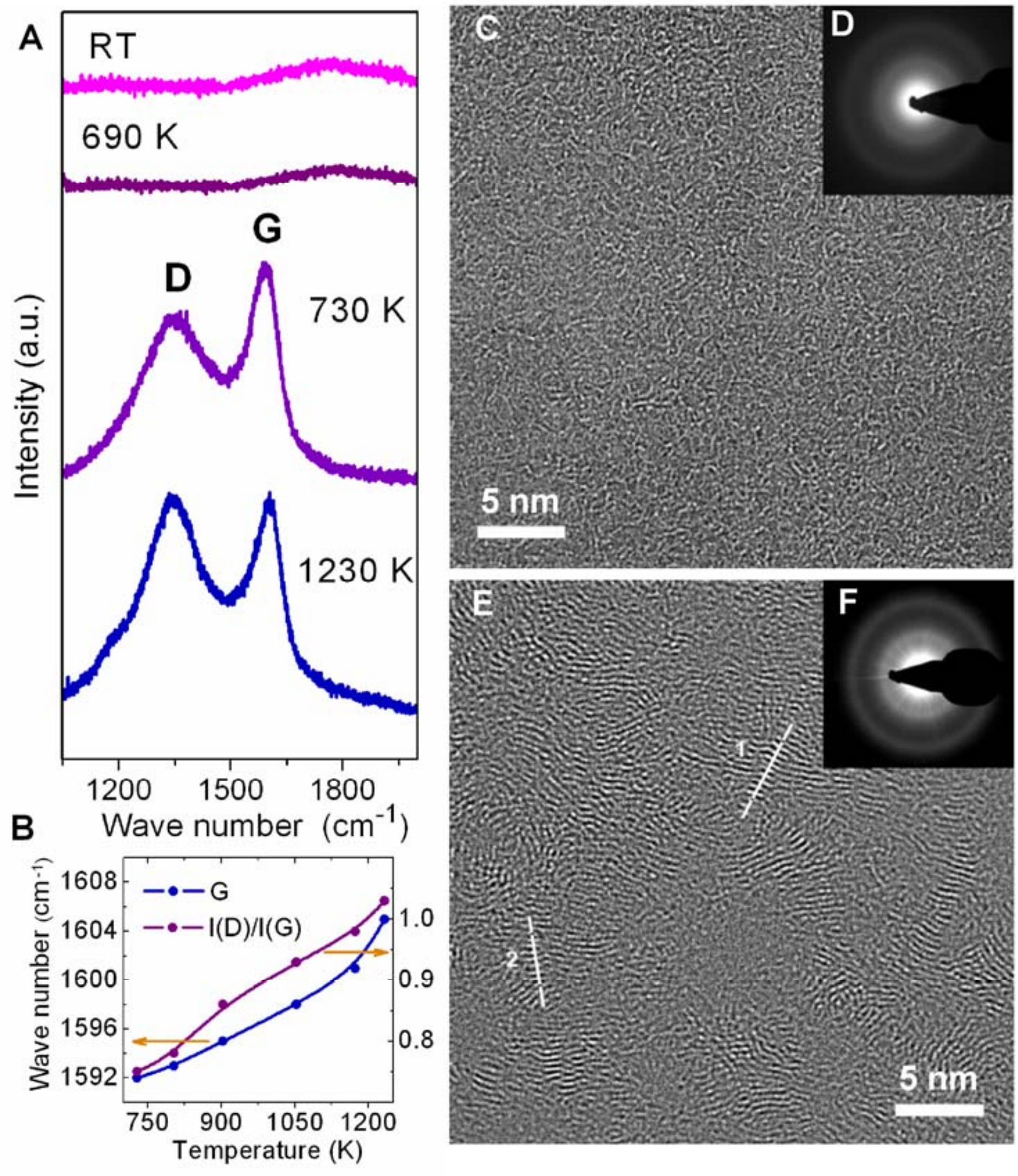

**Fig. 3. Structural transformations of carbon nanosheets upon annealing.** (A), Raman spectra of non-annealed biphenyl nanosheet and nanosheets annealed in vacuum on gold, transferred to silicon oxide substrates. The measurements for different annealing temperatures were performed under ambient conditions. (B), Changes in the position of the G-peak and the intensity ratio of I(D)/I(G) as a function of the annealing temperature. (C), (E) High-resolution phase contrast TEM images and SAED patterns from larger areas of non-annealed (D) and annealed (F) (~1300 K) biphenyl nanosheets. Two typical locations of line profiles (1) and (2), taken for evaluation of the periodicity of the graphitic fringes are depicted in (E). In addition, an enlargement of the diffraction (F) is given in SOM. The rings in (F) can be indexed as belonging to the (0-110) and (1-210) lattice plane spacings of highly in-plane oriented nanocrystalline graphitic sheets.

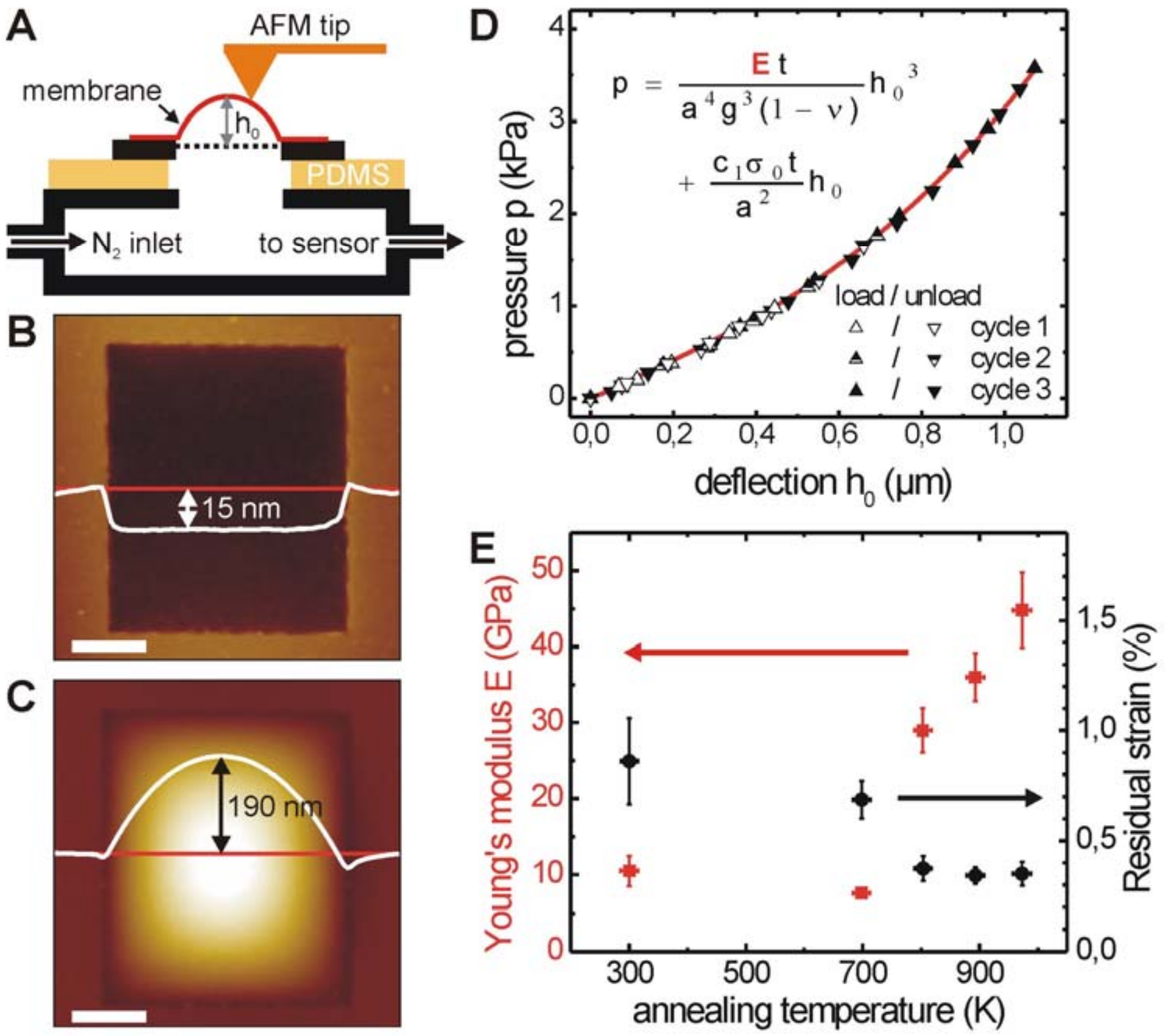

**Fig. 4. Mechanical properties of carbon nanosheets upon annealing.** (A), Schematic representation of the bulging test setup. The home-build pressure cell was mounted into an atomic force microscope (AFM) that measured the membrane deflection. AFM images of a membrane (topography, contact mode) without (B) and with (C) an applied pressure of 450 Pa. Scale bar, 10 µm. Line scans along the red lines are superimposed to the AFM images. (D), The Young's modulus determination is presented for one representative membrane (annealed at ~900 K). First the deflection at the membranes centre is measured for different pressures and then these data are fitted by the displayed dependency which yields the modulus. (E), The Young's modulus as function of annealing temperature. At higher temperatures the modulus shifts towards the value of graphite.

# SUPPORTING INFORMATION

## One nanometer thin carbon nanosheets with tunable conductivity and stiffness

**Andrey Turchanin[1*], André Beyer[1], Christoph Nottbohm[1], Xianghui Zhang[1], Rainer Stosch[2], Alla Sologubenko[3], Joachim Mayer[3], Peter Hinze[2], Thomas Weimann[2], Armin Gölzhäuser[1]**

*[1]Fakultät für Physik, Universität Bielefeld, 33615 Bielefeld, Germany*

*[2]Physikalisch-Technische Bundesanstalt, 38116 Braunschweig, Germany*

*[3]Gemeinschaftslabor für Elektronenmikroskopie, RWTH Aachen, 52074 Aachen, Germany*

E-mail: turchanin@physik.uni-bielefeld.de

# Materials and Methods

## Fabrication and transfer of nanosheets

For the preparation of 1,1'-biphenyl-4-thiol (BPT) SAMs, Fig. S1, we used 300 nm thermally evaporated Au on mica substrates (Georg Albert PVD-Coatings). The substrates were cleaned in a UV/ozone-cleaner (FHR), rinsed with ethanol and blown dry in a stream of nitrogen. They were then immersed in a ~10 mmol solution of BPT in dry, degassed dimethylformamide (DMF) for 72h in a sealed flask under nitrogen. Afterwards samples were rinsed with DMF and ethanol and blown dry with nitrogen. Cross-linking was achieved in high vacuum ($<5*10^{-7}$ mbar) with an electron floodgun (Specs) at an electron energy of 100 eV and typical dose of 50 $mC/cm^2$.

Annealing of the cross-linked nanosheets on Au surfaces was conducted in UHV conditions in Mo sample holders with a resistive heater with the typical heating/cooling rates of ~150 K/h and the annealing time from 0.5 h to 3 h. Annealing temperature was controlled with a Ni/Ni-Cr thermocouple and two-color pyrometer (SensorTherm). Cross-linked biphenylthiol nanosheets on gold films were annealed in vacuum up to ~1200 K, however, the mica substrate is substantially damaged at temperatures above ~1000 K, leading to damage of the gold-film/nanosheet as well. In order to maximize defect free areas, the Au-film was cleaved from the mica by immersion in hydrofluoric acid (48%) for 5 min and transfered it to a clean quartz substrate. Transfer of the nanosheet after annealing follows the procedure as described below.

Transfer of non-annealed and annealed nanosheets was conducted by cleaving the nanosheet from its substrate using a layer of polymethylmethacrylate (PMMA) for stabilization. A ~500 nm thick layer of polymethylmethacrylate (PMMA) was spincoated onto the sample and baked on a hotplate. The Au was cleaved from the mica by immersion in hydrofluoric acid (48%) for 5 min and etched away in an $I_2$/KI-etch bath (~15 min). Afterwards the nanosheet/PMMA was transferred to a $SiO_2$ substrate or TEM grid (Quantifoil, Plano) followed by dissolution of the PMMA in acetone to yield a clean nanosheet. With this method it is possible to obtain freestanding membranes of >100 µm in size. For TEM/SEM/STM measurements nanosheets were also annealed directly on TEM grids using either resistive or e-beam heaters.

**Spectroscopy and Microscopy**

X-ray photoelectron spectra were acquired with an Omicron Multiprobe spectrometer utilizing monochromatic Al $K_\alpha$ radiation under ultra high vacuum (UHV) conditions (~$10^{-10}$ mbar). Binding energies were calibrated with respect to the Au $4f_{7/2}$ peak at 84.0 eV[1], resolution of the spectra corresponds to ~1eV. A constant pass energy mode of the energy analyzer was used. Auger spectroscopy and scanning electron microscopy (SEM) of the suspended membranes were conducted with a scanning Auger microscope (SAM) in connected UHV chamber (Omicron Multiscan). An electron energy of 3 kV and a constant retardation ratio mode of the energy analyser were utilized. Raman spectra were measured with a triple monochromator system Horiba Jobin-Yvon T64000 equipped with a liquid $N_2$-cooled CCD detector and an Olympus BH2 microscope. Data was collected in back-scattering geometry with a spectral resolution of 2 $cm^{-1}$ using 514.5 nm line of an $Ar^+$ laser (as the excitation source). The use of an 80× objective led to a spatial resolution of ~2.5 µm and a laser power on the sample surface of 10 mW. Spectra were calibrated against the 520 $cm^{-1}$ peak of the Si/SiO substrate. Transmission electron microscopy was performed by Philips CM 200 FEG and FEI Titan T operated at 300 keV. Optical images were acquired with an Olympus BX51 microscope with a C5060 camera. AFM was done on an Ntegra system (NT-MDT) in contact mode with cantilevers by NT-MDT (Pt-coated, spring constant 0.1 N/m) as well as Olympus (0.02 and 0.08 N/m).

**Electrical and Mechanical Measurements**

Resistance measurements of suspended nanosheets on a gold grid were conducted in UHV by contacting with the tungsten tip of the scanning tunnelling microscope of an Omicron Multiscan Microscope. The sheet resistance of nanosheets transferred on $SiO_2$ was determined by a four-point measurement using Suess probes and a Keithley SMU Source-Measure Unit (Model 236). Mechanical measurements of nanosheet membranes on silicon window structures were performed under ambient conditions with a home-built pressure cell in a NT-MDT NTEGRA Scanning Probe Microscope. A detailed description of the experimental procedures and data evaluation is given below.

# Supporting Online Text

## Analysis of spectroscopy and microscopy data

X-ray photoelectron spectroscopy (XPS) measurements were conducted as described in the Materials and Methods. The monolayer thickness was calculated by assuming an exponential attenuation of the Au $4f_{7/2}$ (or Si 2p) signals with a photoelectron attenuation length of 36 Å (35 Å)[2]. The XP spectra of C1s, S2p and Auf peaks for pristine, e-beam cross-linked and annealed BPT monolayer on Au are presented in Fig. S2. No other elements have been identified in the widescan XP spectrum. In some pristine BPT samples an O1s signal was seen just at the noise level of the measurement. This signal disappears completely in UHV after electron irradiation and annealing. Electron irradiation and annealing of the samples was conducted in UHV in the spectrometer chamber.

The effective thickness of a pristine monolayer at room temperature is calculated as ~ 1.0 nm. This is in good agreement with the AFM data, Fig. S3. It decreases to a value of ~ 0.7 nm after electron irradiation (50 eV, ~ 50 $mC/cm^2$) and annealing (~ 1000 K), Fig. S2. This correlates with a partial decrease of the C1s intensity and the disappearance of the S2p signal (desorption of sulfur).

The chemical composition of annealed suspended membranes on TEM grids was analysed with a scanning Auger microscope. Fig. S4 presents an Auger spectrum of a BPT nanosheet annealed at ~1300 K on a Quantifoil-on-Mo TEM grid by e-beam heating in vacuum for 5 min. For comparison the Auger spectra of highly oriented pyrolytic graphite (HOPG) and a hole in Quantifoil (space without nanosheet) are presented. Besides the C KLL Auger line no other Auger transitions can be seen in the spectrum of the annealed nanosheet. Thus it consists of only carbon. We observed that the annealing of the nanosheets on Quantifoil-on-Mo TEM grids by e-beam heating at temperatures above 1500 K results in the formation of precipitates in the membranes, Fig. S5a. These are most likely carbides of Mo and W and may result from the hot Mo and W parts of the e-beam heater (W filament, Mo bottom plate of the sample holder). Fig S5b shows an enlargement of the SAED in Fig 3f of the main text. The rings labelled (1) and (2) can be indexed as belonging to the (0-110) and (1-210) lattice plane spacings of highly in-plane oriented nanocrystalline graphitic sheets.

Fig. S6 presents chemical analysis of the transferred nanosheets on oxidized Si wafers. The thickness of the non-annealed BPT nanosheet at room temperature is ~1.2 nm. Some residual contaminations (C, O, F, I) from the fabrication and transfer procedure are observed in the XP spectra. These contaminations disappear after annealing to ~600 K in UHV, correlating with a decrease of the film thickness to ~0.8 nm. Nanosheets heated first on Au and then transferred on oxidized Si did not show any F or I signals even without annealing.

## Analysis of resistivity measurements

### *In situ* resistivity measurements

The resistivity of the monolayer was determined in ultra high vacuum with the aid of a scanning tunneling microscope (STM). For this measurement the nanosheet was prepared on a gold grid (1500 mesh) and contacted by the STM tip in the center of the ~11 µm wide openings. A bias V was then applied between the tip and the gold grid, resulting in a linear current response. The sheet resistivity can be extracted from such measurements by applying a simple model. First it is assumed that a homogenous film is contacted by two electrodes; one electrode, a circular dot, is placed at the centre of a ring-like second electrode that confines the measured area, cf. Fig. S7. The current-voltage-characteristics can be easily calculated for this setup if contact resistances are neglected. Thus, the sheet resistivity $\rho_S$ is determined by

$$\rho_S = \frac{2\pi}{\ln(r_{ring} / r_{dot})} \frac{V}{I}$$

with the inner radius of the ring electrode $r_{ring}$ and the radius of the dot electrode $r_{dot}$. In our case the contact area of the STM tip is not exactly known and the outer electrode is a square-like frame. However, as the ratio of radii enters only as the argument of a logarithmic function, the sheet resistivity is not drastically affected by the electrode geometry. We therefore approximated our setup with this model using a ring electrode radius $r_{ring}$ of 5 μm and a dot electrode radius $r_{dot}$ of 100 nm. Assuming that the true contact area of the tip equals a circle with a radius of 1 Å, our approximation leads to an overestimation of the sheet resistivity by a factor of less than three. This overestimation is not critical in our contribution as we discuss changes in the resistivity of more than five orders of magnitude.

## *Ex situ* resistivity measurements

A four-point probe measurement method was used to determine the sheet resistivity more accurately. The nanosheet was placed on a $SiO_2$ surface. Four probe tips were equidistantly arranged in a line and contacted the film as shown in Fig. S8. A current I was driven through the two outer needles and the voltage drop V between the two inner needles was measured. In this setup the sheet resistance $\rho_S$ is given by[3] :

$$\rho_S = \frac{\pi}{\ln(2)}\frac{V}{I} = 4.532\frac{V}{I}\Omega/\text{square}$$

**Analysis of mechanical measurements**

Bulging tests were employed to measure the Young's modulus of nanosheets transferred onto silicon samples with micron-sized openings. These silicon samples with the suspended nanosheet on top were mounted on a self-made pressure cell, a hollow steel cylinder with two sideway openings for applying and measuring a pressure and one upward opening for connecting with one nanomembrane, cf. Fig. S9a. A layer of polydimethylsiloxane (PDMS) was used to estabilsh a gas-tight seal between the silicon sample and the pressure cell. The nitrogen gas supply and the differential pressure sensor (HCX001D6V, Sensortechnics) were connected with the cell as schematically shown in Fig. S9b. Deflection of the membrane was measured with a NT-MDT NTEGRA Atomic Force Microscope (AFM) in contact mode by employing a platinum-coated silicon cantilever (force constant: 0.1 N/m). The platinum coating reduces the adhesion between the tip and the monolayer. Scans for data acquisition were conducted with a scan-speed of 5 – 8 µm/s and a very low feedback gain of 0.01 – 0.02, whereas scans for imaging used a scan-speed of 15 µm/s and a feedback gain of 0.35. The latter setting yields an improved image quality but is less gentle to the nanomembranes, i.e. the probability of rupture is enhanced.

The deflection setpoint setting adjusts the force that the tip applies to the nanomembrane. This force leads to an indentation as apparent by the 15 nm high step between the silicon frame and the nanomembrane in Fig. 4B of the main text. A systematic variation of the setpoint with the resulting step height is presented in Fig. S10. The linear dependence shows a vanishing step height for a deflection setpoint of zero. This setpoint corresponds to the deflection value of an unperturbed cantilever, e.g. far away from the sample. In other words the cantilever is not bent at the deflection value of zero and therefore it does not apply any force to the nanomembrane at this deflection value. The bulging measurements were performed with a slightly higher setpoint which led to a certain step height $\delta$. This quantity was measured on non-pressurized membranes; it was modeled for arbitrary pressures and it was employed to correct the measured deflection of the nanomembranes as shown in the following. The deflection of the membrane h is given by the AFM height signal $h_{AFM}$ and the step height $\delta$ by:

$$h = h_{AFM} + \delta .$$

Here all quantities are given in reference to the height level of the silicon frame. The step height $\delta$ is defined as positive if the AFM height signal is below the unperturbed membrane deflection. In the example of Fig. 4B the deflection of the membrane h is zero and the step height $\delta$ has a positve value of 15 nm. The step height $\delta$ of non-pressurized membranes was measured and employed to correct the deflection h. However, this measurement was not possible on bulged membranes. Therefore the change of the step height $\delta$ due to the increased tension in a bulged membrane was taken into account by the following calculation[4-6]

$$\delta = \delta_0 \frac{\sigma_0}{\sigma_0 + \frac{2}{3}\frac{E}{1-\nu}\frac{h_C^2}{a^2}}$$

with the step height of non-pressurized membranes of $\delta_0$. All other quantities are explained in the next section. Note, that one approximation of this correction scheme is to assume a constant step height, i.e. $\delta = \delta_0$. This simplification results in an underestimation of the Young's modulus and the residual stress of up to 9 % and 1.6 %, respectively for the presented datasets. Nevertheless, all data were treated with the full correction scheme, only the step height $\delta$ was calculated with the Young's modulus and the residual stress of the simplified scheme.

The change of the membrane deflection due to the applied pressure was employed to determine the Young's modulus and the residual stress. The deflection of the membranes centre $h_C$ is described by[7]

$$p = \frac{Et}{(1-\nu)a^4 g^3(\nu, \frac{b}{a})} h_C^3 + c_1(\tfrac{b}{a}) \frac{\sigma_0 t}{a^2} h_C$$

with the pressure p, the Young's modulus E, the residual stress $\sigma_0$, the membrane thickness t, the Poisson ratio $\nu$ and the length of the membranes short-edge being 2a. The constants g and $c_1$ are taken from the literature[7] and depend only on the membrane's aspect ratio b/a and in the case of g on the Poisson ratio $\nu$. The Young's modulus and the residual stress were determined by fitting the equation above to the measured $p(h_C)$ data. Note that each fitting constant is a measure for just one quantity: the Young's modulus E or the residual stress $\sigma_0$. The dimension of the membrane was measured in scanning electron micrographs (SEM) which allowed a higher precession as the values extracted from AFM images. In the case of Fig. 4B-D the half length of the membranes short edge a was determined to 15.93 µm. All calculations were carried out with a thickness t of 1 nm and a Poisson ratio $\nu$ of 0.35. The latter value is not known, so a typical value for polymers was used.

The residual strain $\varepsilon_0$ was calculated from the residual stress and the Young's modulus by[5]

$$\varepsilon_0 = \frac{(1-\nu)\sigma_0}{E}.$$

The strain of the bulged membranes was determined from the membrane deflection $h_C$ by[6]

$$\varepsilon = \varepsilon_0 + \frac{2{h_C}^2}{3a^2}.$$

In all presented bulging measurements the strain did not exceed a value of 1.3 %.

The accuracy of this Young's modulus measurement will be discussed in two parts, first in terms of the comparability of the presented data then in terms of the absolute accuracy. The most striking point in this error analysis is the strong effect of an uncertainty in the membrane dimension due to its contribution as fourth power to the Young's modulus. Therefore any relative measurement error in the membrane size results in a fourfold contribution to the uncertainty of the Young's modulus. This and contributions from the fitting quality and the constant g were determined for each measurement and are given as error bars in Fig. 4 of the main text. Additional sources of uncertainty lead to a constant but unknown correction factor for all measurements. Therefore these error contributions are discussed separately in the following. All uncertainties of the Young's modulus related to the calibration of the AFM piezo, the scale bar in the SEM and the pressure sensor sum up to 15 %. The accuracy of the constant g, given in Ref. [7], was estimated from the comparison with experimental data in Ref. [7] to 3 % which gives a contribution to the Young's modulus uncertainty of 9 %. The monolayer thickness uncertainty was estimated to 15 %. Thus these error contributions add up to about 40 %.

# Supporting Figures and Legends

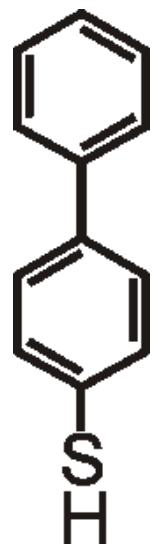

**Fig. S1** 1,1'-biphenyl-4-thiol (BPT)

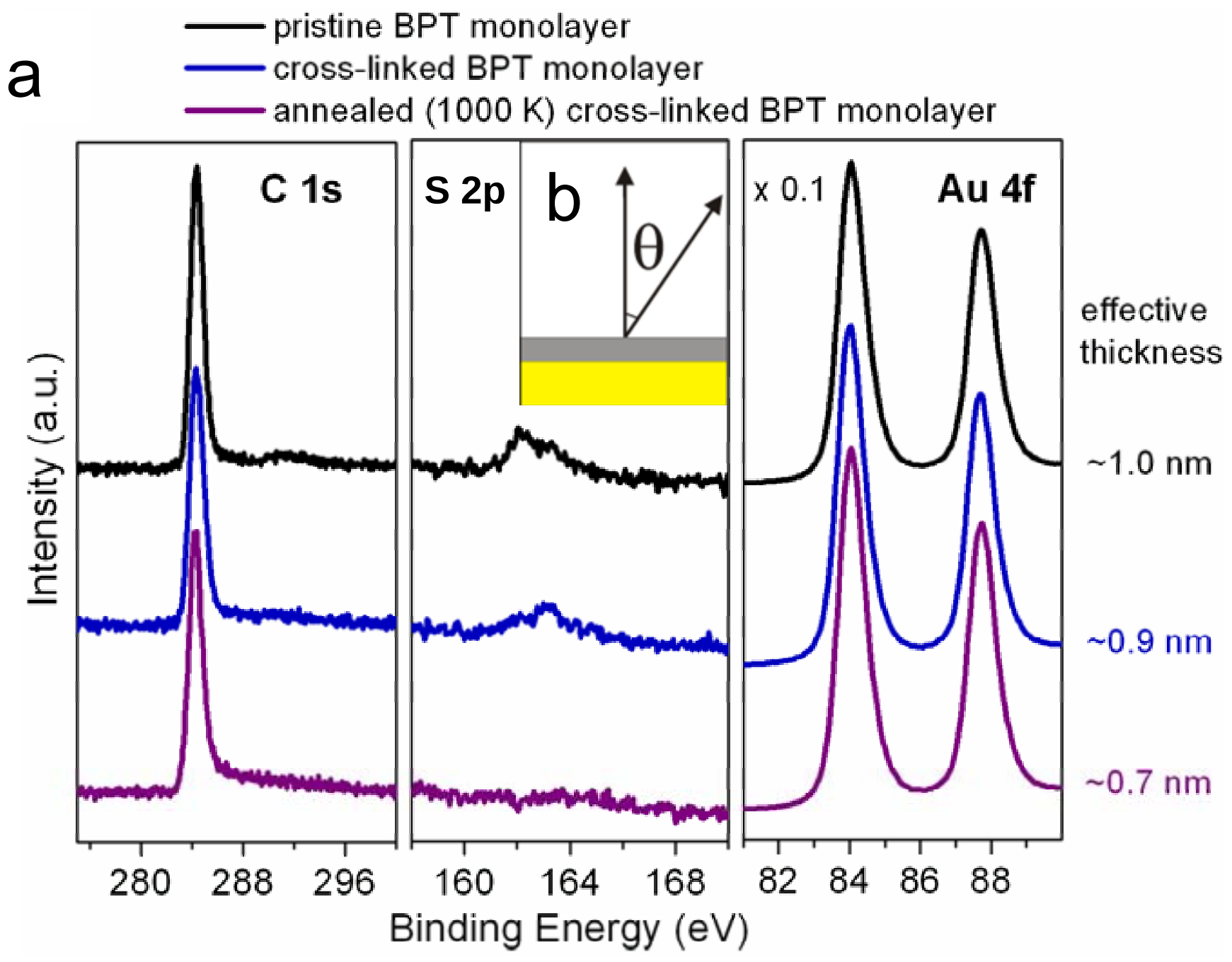

**Fig. S2** XPS characterization of BPT nanosheets on Au surface. (**a**) XP spectra of pristine, electron irradiated and annealed BPT samples acquired with a monochromatic Al-$K_{\alpha}$ source at a detection angle of the electron analyzer of 18° (monochromatic Al $K_{\alpha}$ radiation). (**b**) Schematic representation of the detection angle in the XPS measurements.

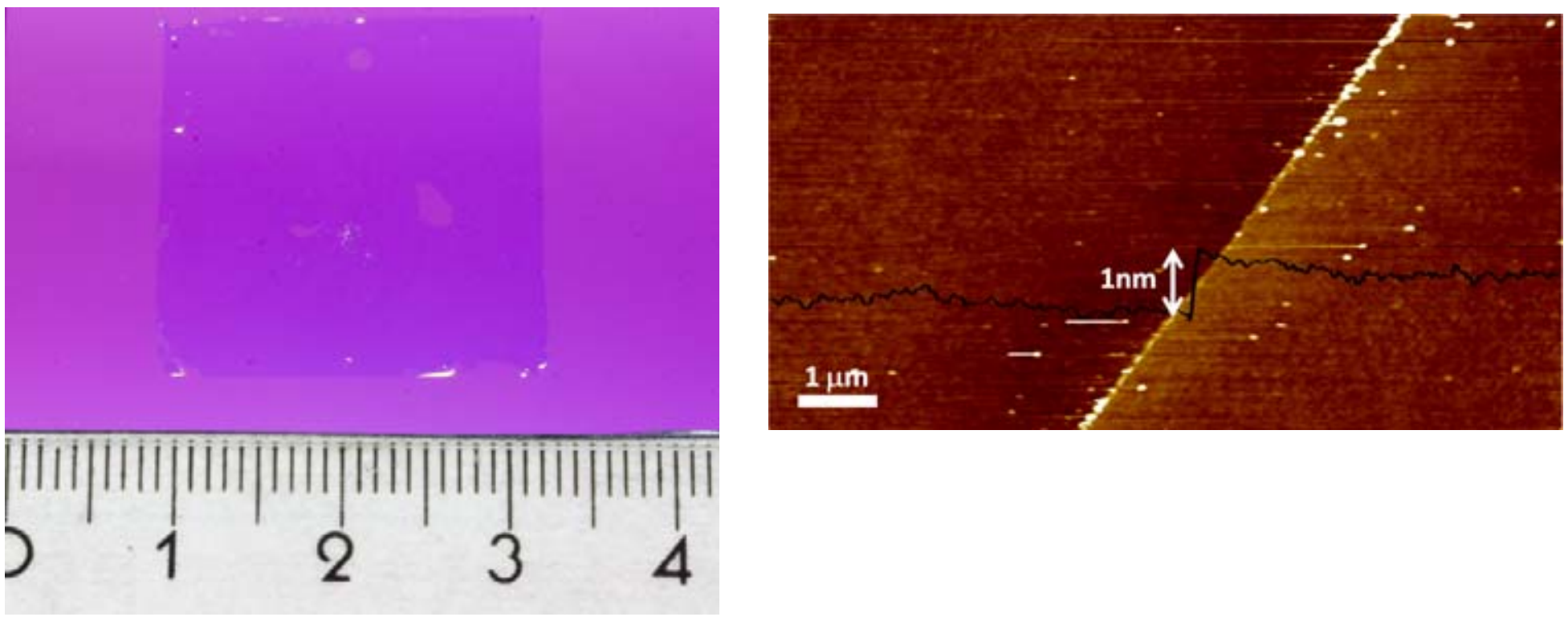

**Fig. S3** Optical micrograph of an extremely large (~5 cm$^2$) cross-linked BPT nanosheet transferred on a 300 nm silicon oxide layer on silicon (left). AFM micrograph of the nanosheet edge showing its thickness of ~1nm (right).

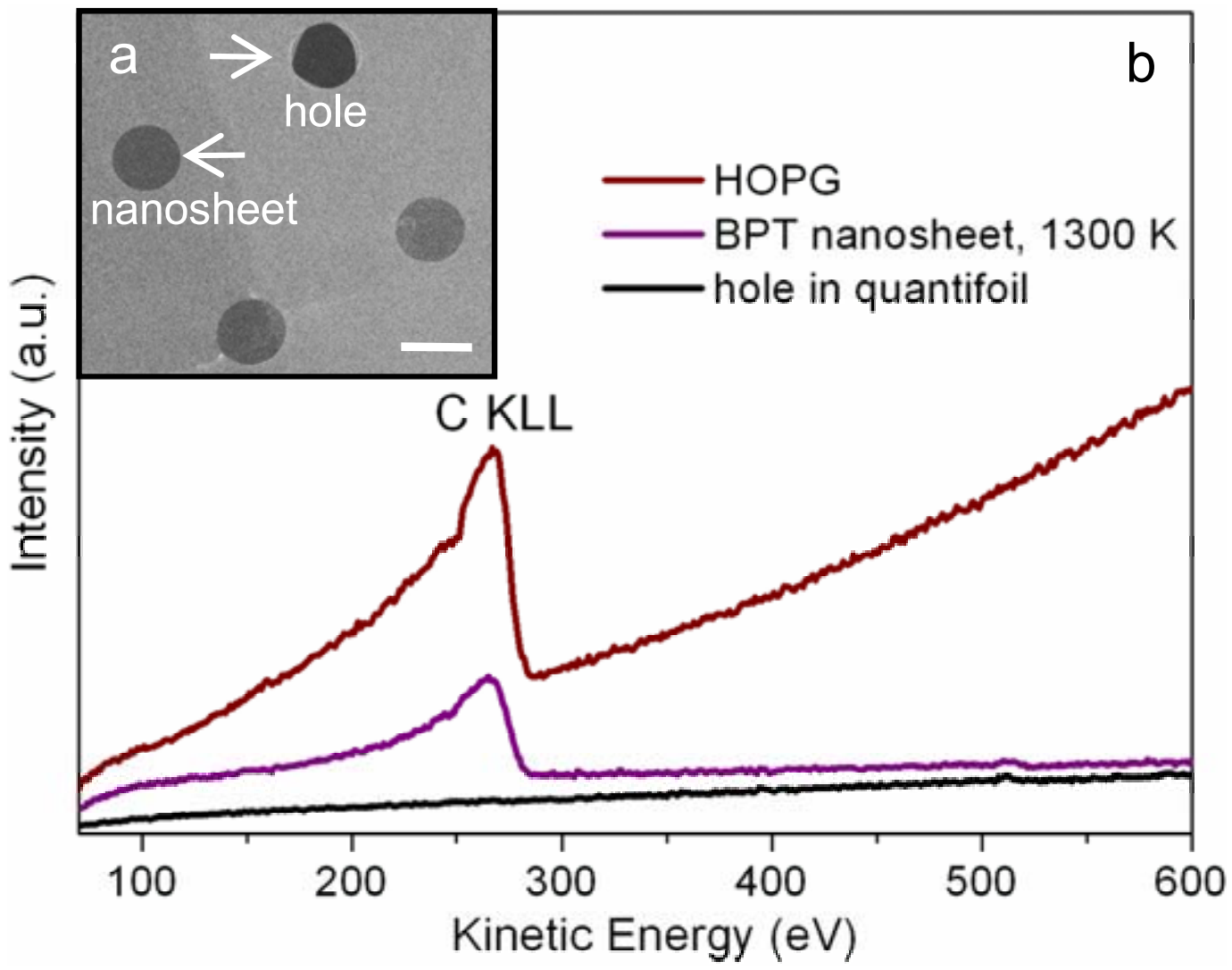

**Fig. S4** Auger microscopy characterization of annealed suspended nanosheets. (**a**) SEM Image of the analyzed area. (**b**) Auger spectra of a BPT nanosheet annealed for 5 min on a Quantifoil-on-Mo TEM grid in vacuum and HOPG surface. Excitation energy 3 kV.

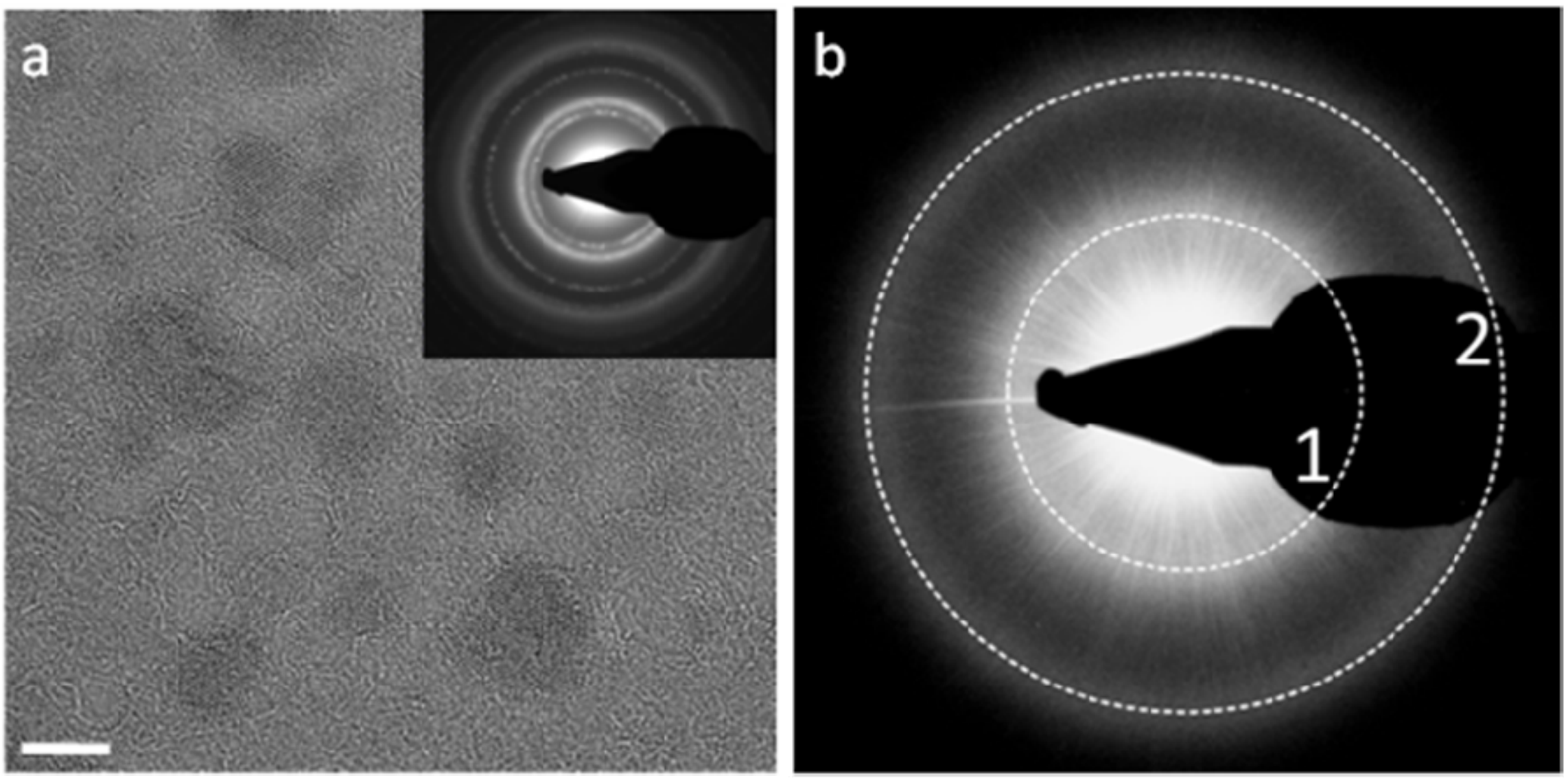

**Fig. S5** a) HRTEM micrograph of a BPT nanosheet annealed 5 min in vacuum on a Quantifoil-on-Mo TEM grid at ~1700 K. Scale bar, 5 nm. Inset: SAED of somewhat larger area. b) Enlarged SAED from Fig 3f (main text) The rings labelled (1) and (2) can be indexed as belonging to the (0-110) and (1-210) lattice plane spacings of highly in-plane oriented nanocrystalline graphitic sheets.

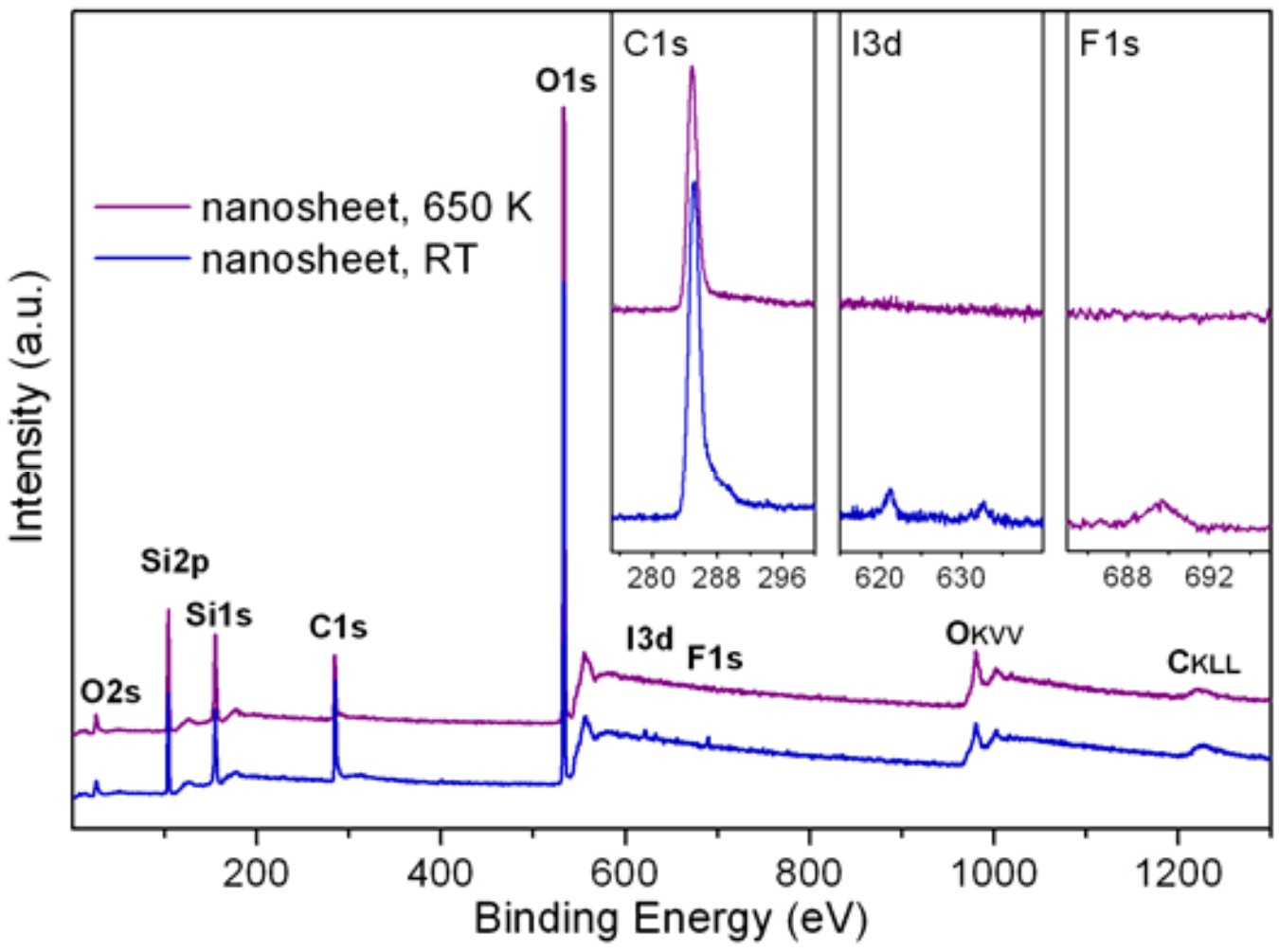

**Fig. S6** XPS of a BPT nanosheet transferred to an oxidized Si wafer before and after annealing at ~ 650 K. Monochromatic Al-$K_\alpha$ radiation, angle of the electron analyzer of 18° relative to the surface normal.

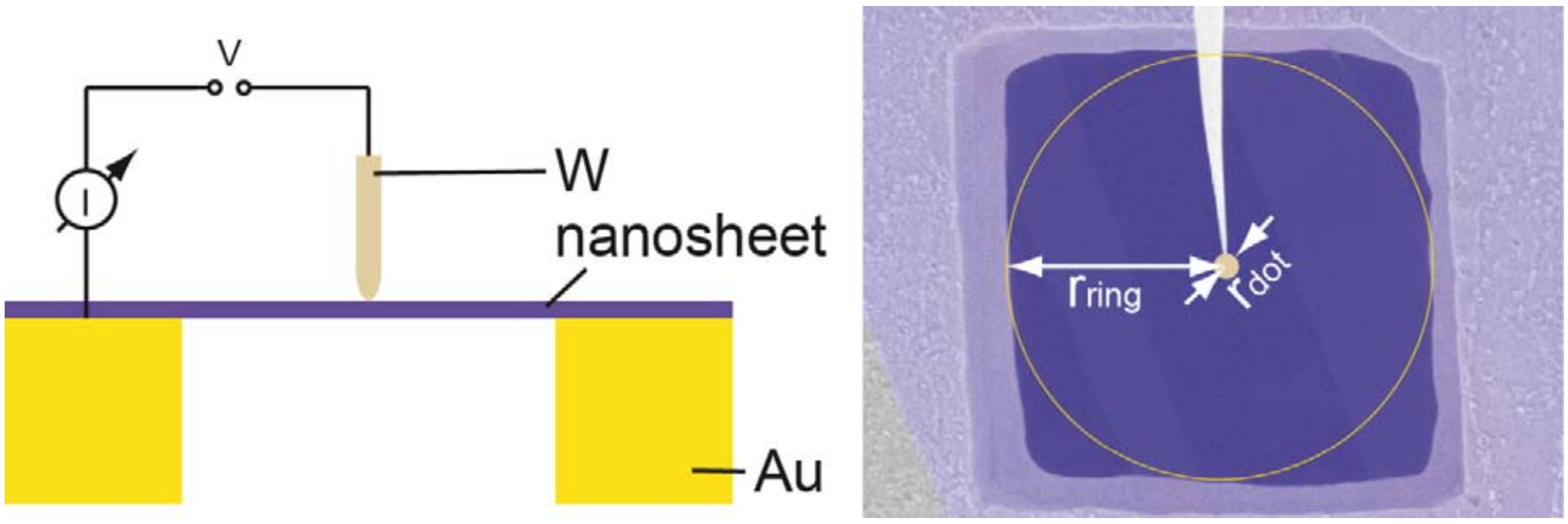

**Fig. S7** Schematic drawing of the 2-probe measurement (left) and SEM image (right) including the radii of the model that was used to determine the sheet resistivity. The nanosheet is contacted by a tungsten tip of an STM and a voltage is applied between the tip and the supporting gold grid to measure the current response.

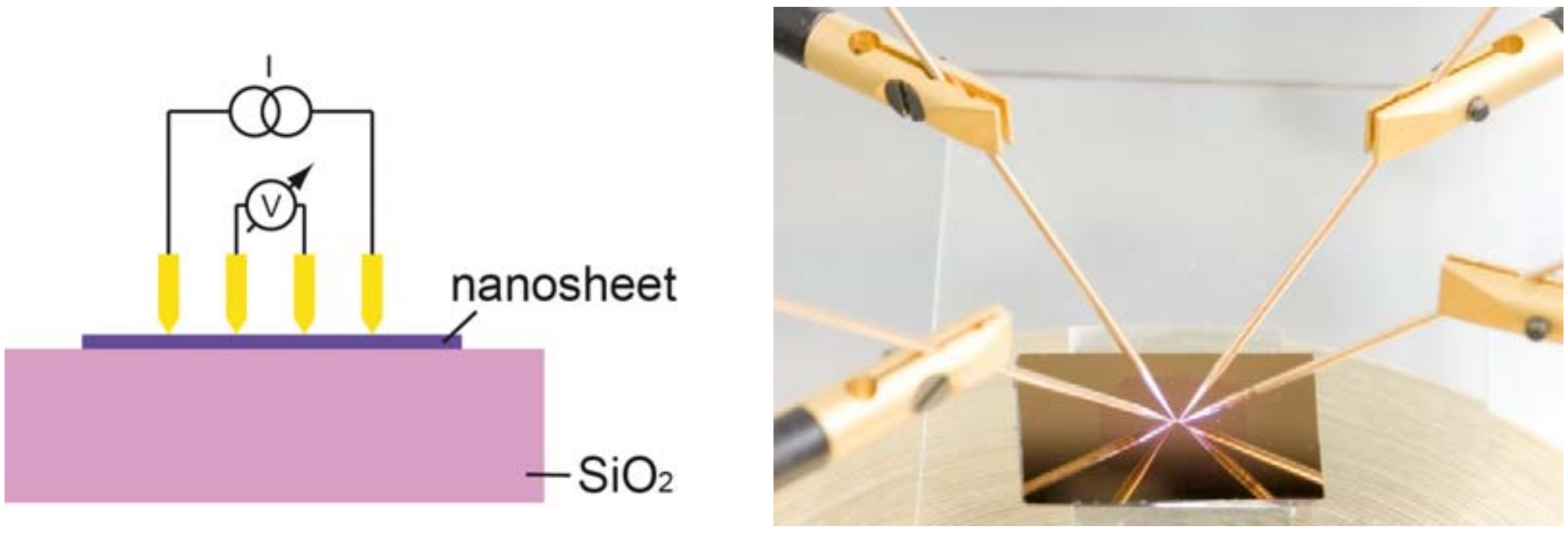

**Fig. S8** Schematic drawing of the 4-probe measurement (left) and photograph of the actual setup (right). Four probes are equidistantly brought into contact with the nanosheet; a current is fed through the outer probes and the voltage drop is measured at the inner two probes.

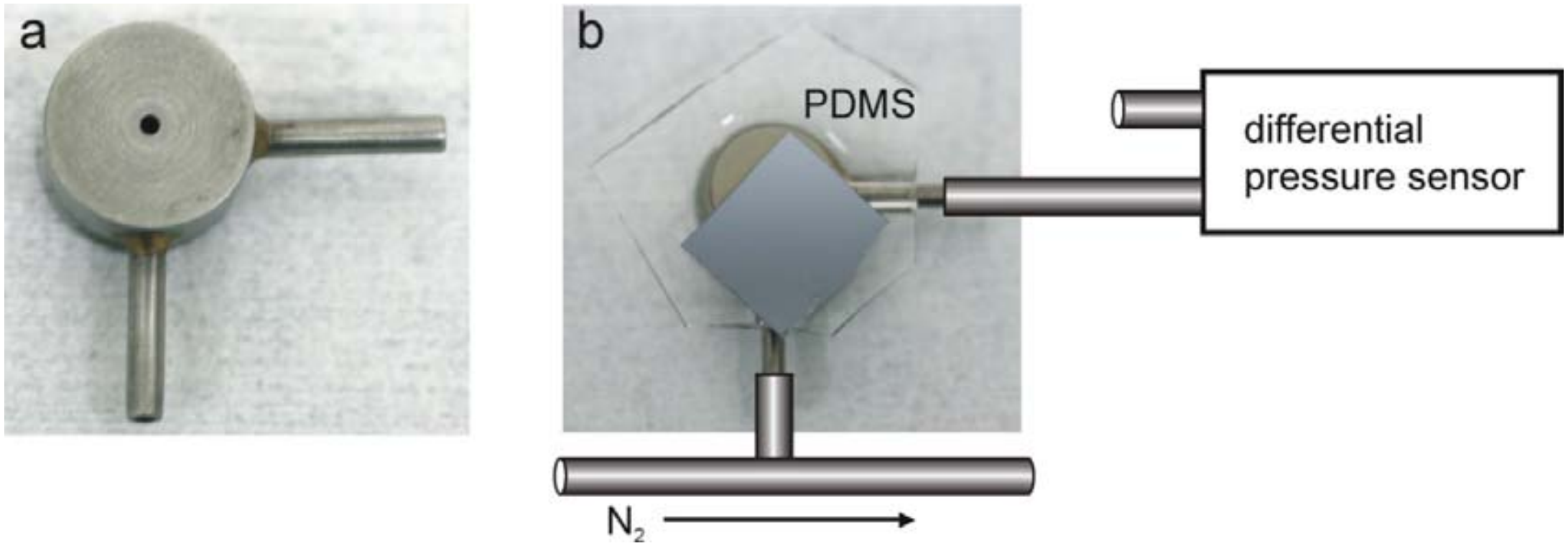

**Fig. S9** Photograph of the pressure cell. **b** Schematic of the bulging test setup and photograph of the pressure cell with one sample mounted.

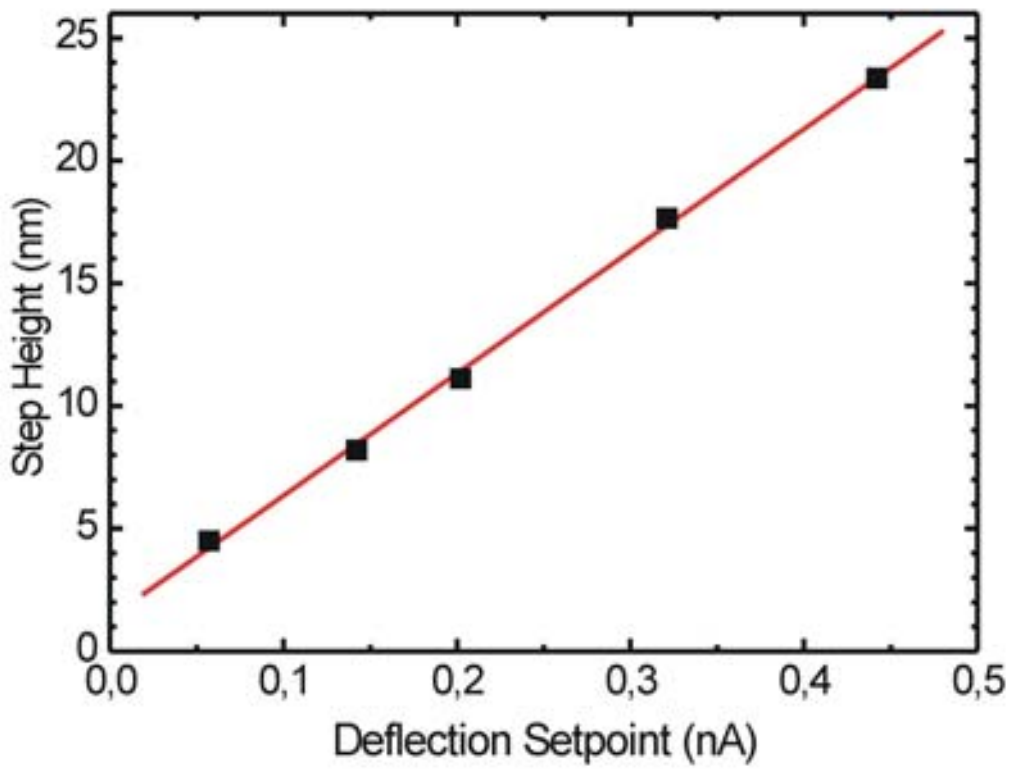

**Fig. S10** AFM height step at the border between the nanomembrane and the silicon frame as function of the deflection setpoint. A zero setpoint corresponds to the deflection of an unperturbed cantilever, e.g. far away from the sample.

## Supporting References and Notes